%% file: econ_v17.tex
\documentclass[11pt]{amsart}

\usepackage{amsmath, amssymb, amsfonts, amsthm}
\usepackage{graphicx}
\usepackage{xcolor}
\usepackage{listings}
\usepackage{needspace}
\usepackage{natbib}
\usepackage[hidelinks]{hyperref}
\usepackage{url}
\usepackage{booktabs}
\usepackage[font=small,labelsep=period]{caption}
\usepackage{inconsolata}
\usepackage[margin=1in]{geometry}

\newcommand{\Bias}{\operatorname{Bias}}

\DeclareMathOperator{\Var}{Var}

\newcommand{\bE}{\mathbf{E}}         
\newcommand{\bR}{\mathbf{R}}  
\newcommand{\bI}{\mathbf{I}}          
\newcommand{\bk}{\mathbf{k}}

\newcommand{\bse}{\boldsymbol{\eta}}
\newcommand{\bsn}{\boldsymbol{\nu_k}}
\newcommand{\bZ}{{\mathbf Z}}
\newcommand{\bM}{{\mathbf M}}
\newcommand{\bV}{{\mathbf V}}
\newcommand{\by}{{\mathbf y}}
\newcommand{\bsG}{{\mathbf G}}
\newcommand{\bG}{{\mathbf G}}

\newcommand{\bX}{{\mathbf X}}

\newcommand{\bsB}{\boldsymbol{\beta}}
\newcommand{\bds}[1]{\boldsymbol{#1}}

\def\spacingset#1{\renewcommand{\baselinestretch}{#1}\small\normalsize}
\spacingset{1.5}

\title{Parameter-Specific Bias Diagnostics in Random-Effects Panel Data Models}

\author{Andrew T. Karl}
\date{17 November 2026}
\begin{document}

\begin{abstract}
The Hausman specification test assesses the random-effects specification by comparing random-effects and fixed-effects estimators. This note uses the diagnostic of Karl and Zimmerman (2021) to give a coefficient-specific sensitivity summary from one fitted mixed model. The diagnostic measures alignment between fitted random effects and the weights used to estimate a coefficient or contrast. Its value need not equal finite-sample bias: fitted effects are shrunken, and response-dependent covariance estimates introduce an additional weighted-error term. We explain this distinction, describe simulations that refit the covariance parameters and retain the latent-effect and weighted-error contributions, and illustrate the calculations with gasoline-demand data and a multiple-membership value-added model. Connecting the diagnostic to established generalized least squares sensitivity analysis, including Vasnev's (2010) balanced-panel results, we identify it as the negative derivative of a coefficient or contrast with respect to the log scale of the random-effects covariance. With fitted covariance shapes held fixed, integrating the diagnostic over this log scale, upward from the fitted model, gives the random-effects estimate minus the fixed-effects estimate for a contrast estimable under both models. We also discuss conditioning on stochastic regressors and the additional assumptions needed for calibrated inference.
\end{abstract}

\keywords{linear mixed model, Hausman test, correlated random effects, permutation test}

\maketitle

\section*{Use of generative AI}
GPT6 Astra (OpenAI) wrote Section~\ref{sec:future} in its entirety. It also
assisted with review and rewriting of the rest of the manuscript, including
mathematical checks, prose revisions, and supporting code.
Claude Opus 5.5 (Anthropic) provided an independent review of the manuscript
and software.
The author is responsible for the final content and interpretation.

\section{Introduction}
\label{sec:intro}

Consider a linear mixed model:
\begin{equation}\label{eq:mixedmodel}
\by = \bX\bsB + \bZ\bse + \bds{\epsilon}.
\end{equation}
Here $\bse \sim N(\bds{0},\bsG)$ is independent of $\bds{\epsilon} \sim N(\bds{0},\bR)$. The response $\by$ is an $n \times 1$ vector, $\bX$ is an $n \times p$ fixed-effects design
matrix with associated coefficient vector $\bsB$, and $\bZ$ is an $n \times m$
random-effects design matrix with random-effects vector $\bse$. The matrices
$\bsG$ and $\bR$ denote the $m \times m$ random-effects covariance matrix and the
$n \times n$ error covariance matrix, respectively.

In random-effects (RE) panel-data applications, consistency of the RE estimator requires the
usual orthogonality and regularity conditions. A common sufficient exogeneity
condition is that the individual effects have conditional mean zero given the
regressors; this is stronger than zero correlation. See, for example,
\citet{hausman2} and \citet{wooldridge}. The Hausman test
\citep{hausman} is the classical tool for assessing this specification by comparing the
RE and fixed-effects (FE) estimators. In its original formulation, this
comparison involves re-fitting the model with $\bse$ treated as an additional vector of
fixed effects, although correlated random-effects (CRE) implementations in the spirit of
\citet{mundlak} and \citet{wooldridge} deliver Hausman-type specification tests from a
single augmented RE fit; see Section~\ref{sec:hausman_bias_tests}. Under the null
hypothesis, both RE and FE are consistent, so their estimates should differ only by
sampling variation. Under alternatives involving dependence between regressors and individual effects,
the FE estimator remains consistent if strict exogeneity of the idiosyncratic errors
and the other FE regularity conditions continue to hold. A
large discrepancy is therefore evidence against the RE specification
\citep{hausman,hausman2}.

The Hausman test concerns consistency: it asks whether the RE estimator converges to the
true parameter values as the sample size grows. Bias, by contrast, is the systematic
difference between an estimator's expectation and the target parameter at a given sample
size. These are distinct concepts -- an estimator may be asymptotically consistent yet
still exhibit finite-sample bias -- but similar forms of misspecification, especially
failures of exogeneity, can affect both. It is therefore natural to use a bias
diagnostic alongside the Hausman test when assessing possible misspecification. In what
follows, we keep these roles separate: the Hausman test serves as a classical large-sample
specification check for the RE estimator, whereas the diagnostic of \citet{kz} is used as
an internal tool to summarize coefficient-specific alignment within the same RE fit.
Its units and sign are informative about sensitivity, but its numerical value need
not equal the estimator's finite-sample bias. Coefficient-specific Hausman-type
comparisons are also available \citep{frondel2010}; the internal diagnostic
is a different descriptive summary.

We draw on the work of \citet{kz}, who develop the diagnostic
$\hat{\bsn}'\hat{\bse}$ for linear mixed models with a stochastic random-effects
design matrix. It can be computed for a coefficient or contrast from one fitted
model and compared with permutations of fitted random effects. That feature is
useful when constructing a separate FE or CRE fit would require substantial
additional modeling work. The corrected author manuscript of \citet{kz}, associated with
\href{https://arxiv.org/abs/2003.08087v3}{arXiv:2003.08087v3}, adds a weighted-error
contribution that the published bias identity omitted and that arises when
covariance parameters are estimated from the response.
Versions of this note before October 2026 also omitted that term. We distinguish the internal diagnostic, a permutation
reference, and a simulation-based assessment under a specified data-generating
mechanism.

Sensitivity of generalized least squares (GLS) estimates to covariance
assumptions has an established literature
\citep{banerjeemagnus1999,magnusvasnev2007}. For balanced random-intercept
panels, \citet{vasnev2010} derives GLS sensitivity and its mean and variance
at the true variance-ratio parameter; under normal effects and errors, the
sensitivity is also normally distributed. The same paper shows that the joint sensitivity
statistic equals the Hausman statistic when the variance estimates are matched,
and places the within, pooled least-squares and feasible GLS estimators on one
curve. We connect this
sensitivity approach to the diagnostic of \citet{kz}, distinguishing local
movement of an estimate from its accumulated movement along a covariance-scale
path. The purpose is to supplement specification tests with an interpretable
summary from one fit, while separating these algebraic connections from the
assumptions required for calibrated inference.

\section{Hausman and Bias Diagnostics}
\label{sec:hausman_bias_tests}

\subsection{The Hausman Test}
The Hausman test \citep{hausman} compares the FE and RE estimators for a common
vector of estimable coefficients, denoted by $\bsB$ in this comparison,
$\hat{\bsB}_{\text{FE}}$ and $\hat{\bsB}_{\text{RE}}$. Under the null hypothesis, both
estimators are consistent and RE is asymptotically efficient within the comparison.
The covariance matrices below must be estimated consistently under that null.
Separately estimated covariance matrices can yield an indefinite difference;
compatible covariance estimation is therefore important \citep{legallo2023}. Writing $\bM^-$ for a generalized inverse of a matrix
$\bM$, the test statistic is
\[
H = (\hat{\bsB}_{\text{RE}}-\hat{\bsB}_{\text{FE}})'
    \bigl[\Var(\hat{\bsB}_{\text{FE}})-\Var(\hat{\bsB}_{\text{RE}})\bigr]^{-}
    (\hat{\bsB}_{\text{RE}}-\hat{\bsB}_{\text{FE}}).
\]
This statistic has an asymptotic $\chi^2(r)$ null distribution under the usual Hausman regularity
conditions, where $r$ is the rank of
$\Var(\hat{\bsB}_{\text{FE}})-\Var(\hat{\bsB}_{\text{RE}})$ \citep{hausman3}. A
significant value of $H$ provides evidence against the RE orthogonality assumptions and,
under the usual conditions, against consistency of the RE estimator. In the canonical
panel-data setting, this statistic is derived under homoskedastic, serially
uncorrelated idiosyncratic errors, that is, $\bR = \sigma^2 \bI_n$. In the linear mixed
model formulation we write the model with a general error covariance matrix $\bR$; the
gasoline example below maintains the standard homoskedastic assumption, whereas the
value-added model in Section~\ref{sec:vam} uses a block-diagonal $\bR$.

\subsection{The internal diagnostic and the full bias decomposition}
Let $\hat{\bG}$ and $\hat{\bR}$ be covariance estimates, and write
$\hat{\bV}=\bZ\hat{\bG}\bZ'+\hat{\bR}$. Assume $\hat{\bV}$ is positive definite
and the contrast $\bk'\bsB$ is estimable for every design under consideration.
With a generalized inverse when needed, define
\begin{equation}\label{eq:beta_hat}
\hat{\bM}=(\bX'\hat{\bV}^{-1}\bX)^{-}\bX'\hat{\bV}^{-1},
\qquad \hat{\bsB}_{\mathrm{RE}}=\hat{\bM}\by.
\end{equation}
For a $p\times1$ contrast vector $\bk$, let
\begin{equation}\label{eq:n_hat}
\hat{\bsn}'=\bk'\hat{\bM}\bZ,
\qquad D_k=\hat{\bsn}'\hat{\bse}.
\end{equation}
The predictor $\hat{\bse}$ is computed under the working mixed model. Its usual
description as an empirical best linear unbiased predictor does not
guarantee unbiased prediction under dependence
between the design and the random effects.

Substitution of \eqref{eq:mixedmodel}, using $\bk'\hat{\bM}\bX=\bk'$, gives the
realized estimation error
\begin{equation}\label{eq:error}
\bk'\hat{\bsB}_{\mathrm{RE}}-\bk'\bsB
=\hat{\bsn}'\bse+\bk'\hat{\bM}\bds{\epsilon}.
\end{equation}
Thus, whenever the expectations exist, the corrected bias identity is
\begin{equation}\label{eq:bias}
\Bias(\bk'\hat{\bsB}_{\mathrm{RE}})
=\bE[\hat{\bsn}'\bse]+C_k,
\qquad C_k=\bE[\bk'\hat{\bM}\bds{\epsilon}].
\end{equation}
For now these expectations use the fixed-$\bX$, stochastic-$\bZ$ framework of
\citet{kz}; Section~\ref{sec:future} also considers stochastic $\bX$. If the
covariance weights are specified functions of the design and
$\bE(\bds{\epsilon}\mid\bX,\bZ)=0$, then $C_k=0$. Covariance estimates obtained
from $\by$ generally depend on the same errors, so this argument does not apply
to fitted weights. Independence of the errors from the design alone does not
remove $C_k$.

The diagnostic $D_k$ substitutes fitted random effects for the latent effects
in the first contribution. It neither estimates $C_k$ nor, in general, estimates
$\bE[\hat{\bsn}'\bse]$ without shrinkage or plug-in error. A small $D_k$ therefore
does not establish small total bias, and subtracting it from a coefficient is
not a generally justified bias correction.

Some familiar zero-bias cases require care. With specified residual covariance
shape $\bR_0$ and $\bX'\bR_0^{-1}\bZ=0$, the GLS coefficient weights reduce to
those from $\bR_0$ alone, even when the random-effects covariance is fitted.
Under conditional mean-zero errors, both contributions in \eqref{eq:bias}
then vanish. This conclusion need not hold if a fitted residual covariance
changes the shape of $\bR_0$. Under the standard centered Gaussian model,
joint independence of $(\bse,\bds{\epsilon})$ from the design and even, translation-invariant
covariance estimation also yield unbiasedness by symmetry, provided the
expectation exists \citep{kackar81}; independence of $\bse$ and $\bZ$ alone
is insufficient. Finally, an FE estimator whose weights
annihilate $\bZ$ is conditionally unbiased when its contrast is estimable, its
residual weights are specified given the design, and
$\bE(\bds{\epsilon}\mid\bX,\bZ)=0$.

\subsection{Permutation references and response-refit simulations}
The permutation diagnostic holds $\hat{\bsn}$ fixed and compares $D_k$ with
$\hat{\bsn}'\bds{\pi}(\hat{\bse})$, permuting only among compatible random-effect
levels or blocks. This gives a reference for the observed alignment. Even
exchangeable latent effects need not have exchangeable predictions: unequal
information and shrinkage can change the distribution of individual predictions.
Consequently, these fitted-effect permutations do not in general provide exact tests of zero
coefficient bias or of RE exogeneity. The packages used here report the
two-sided Monte Carlo tail fraction
\begin{equation}\label{eq:permutation}
p_{\mathrm{perm}}=\frac{1+\sum_{b=1}^{B}
\mathbf{1}\{|D_k^{\pi_b}|\ge |D_k|\}}{B+1}.
\end{equation}
The added one avoids reporting zero from finitely many sampled permutations
\citep{phipson2010}; it does not supply the missing exchangeability assumption.
Signed diagnostics and permutation percentiles can be useful even when these
tail fractions are interpreted descriptively.

A different calculation simulates responses and refits all covariance parameters.
For example, fix the observed design and choose generating values
$\bsB_0$, $\bse_0$, and $\bR_0$. For each independent draw
$\bds{\epsilon}^{(b)}\sim N(0,\bR_0)$, generate
$\by^{(b)}=\bX\bsB_0+\bZ\bse_0+\bds{\epsilon}^{(b)}$ and refit the model.
Retain the two components
\begin{equation}\label{eq:simulation}
A_k^{(b)}=\bk'\hat{\bM}^{(b)}\bZ\bse_0,
\qquad E_k^{(b)}=\bk'\hat{\bM}^{(b)}\bds{\epsilon}^{(b)}.
\end{equation}
Their sum is exactly the simulated coefficient error. Their averages estimate
the two contributions to bias under this conditional generating scenario,
with Monte Carlo standard errors computed from the independent replicates.
Such conditional bias can be nonzero even under a correctly specified RE
model: unconditional unbiasedness averages over the random effects as well
as the errors.
To improve precision for the weighted-error mean, one can also average
$\bk'(\hat{\bM}^{(b)}-\bM_0)\bds{\epsilon}^{(b)}$, where $\bM_0$ is a fixed
reference GLS matrix. The subtracted term has mean zero; this control variate
has the same target, although variance reduction is not guaranteed.

The \texttt{simulate\_bias} functions in \texttt{mixedbiastest}
and \texttt{GPvam} \citep{mixedbiastest,gpvam} implement this calculation
for their supported model classes, by default holding the original fitted
random effects fixed and drawing errors from the fitted residual model.
These generating effects retain their prediction shrinkage; uncertainty in
the chosen generating values is not propagated. The calculation answers a
conditional sensitivity question. It does not recover the actual distribution
of random effects given the design. To study bias induced by stochastic
regressors or assignment, a fuller simulation must specify and regenerate their
joint distribution with the latent effects, then refit the entire procedure.
A simulation that independently redraws random effects under a standard RE null
can assess that null model; it cannot validate the absence of endogenous
assignment in the observed data.

\subsection{Relation to correlated random effects and Mundlak--Wooldridge tests}

In the panel-data literature, a common way to test the random-effects specification is
through correlated random effects (CRE) or ``Mundlak'' type augmentations of the
RE model. Suppose we have the panel-data setup
\[
y_{it} = x_{it}'\beta + c_i + \varepsilon_{it}.
\]
Here $c_i$ is the individual effect and $\varepsilon_{it}$ is the idiosyncratic error. Under the usual
random-effects assumptions, including strict exogeneity and
$E[c_i \mid x_{i1},\ldots,x_{iT_i}] = 0$,
the random-effects estimator is consistent. \citet{mundlak} models potential
correlation between $x_{it}$ and $c_i$ by including the group means $\bar x_i$ of the time-varying regressors in the
regression. \citet{wooldridge} shows that a simple Hausman-type alternative is to
estimate the augmented random-effects model
\[
y_{it} = x_{it}'\beta + \bar x_i'\gamma + u_i + \varepsilon_{it}.
\]
Here $u_i$ is the remaining individual effect after the conditional-mean
augmentation. One can apply a (robust) Wald test of $H_0:\gamma=0$. In a balanced linear panel with the same regressors and the usual null covariance
assumptions, this CRE comparison gives a Hausman-type test from one augmented fit
\citep{mundlak,wooldridge}. Robust implementations based on covariance estimators that are
consistent under general forms of heteroskedasticity and serial correlation are widely
used in applied work; see, e.g., \citet{wooldridge}.

In terms of the linear mixed model~\eqref{eq:mixedmodel}, stacking the panel outcomes
$\{y_{it}\}$ into $\by$ and the regressors $\{x_{it}\}$ into the rows of $\bX$ yields
$\by = \bX\bsB + \bZ\bse + \bds{\epsilon}$, where $\bse$ collects the individual effects
$c_i$ and $\bZ$ is the $n \times N$ incidence matrix (with $N$ the number of individuals)
that has a one in the row corresponding to observation $(i,t)$ and column $i$ when that
observation belongs to individual $i$. In a conventional random-intercept panel, this
incidence matrix is usually treated as fixed once the sample has been defined. The
classical RE exogeneity condition then concerns dependence between the random effects
$c_i$ and the regressors $x_{it}$, whereas the formal stochastic-$\bZ$ framework of
\citet{kz} concerns dependence between $\bse$ and the assignment structure encoded by
$\bZ$. The two setups are therefore related but not identical.

For standard panels, we view $\hat{\bsn}'\hat{\bse}$ primarily as a descriptive,
coefficient-specific summary from the fitted RE model rather than as a literal test of a
stochastic-$\bZ$ assignment mechanism. The diagnostic depends on the fitted RE weights for the coefficient or
contrast of interest. Section~\ref{sec:future} relates it to local movement
along a covariance-scale path; the full RE-minus-FE difference accumulates
movement over that path. In applications where the assignment structure encoded by $\bZ$ is itself
plausibly stochastic -- for example, multiple-membership or assignment problems -- the
interpretation developed by \citet{kz} applies more directly. In either case, the
diagnostic is best viewed as complementary to, rather than a substitute for, the
Hausman or Mundlak--Wooldridge tests: it does not test the global null
$E[c_i \mid x_{i1},\ldots,x_{iT_i}] = 0$, but instead supplies a parameter-specific
finite-sample summary from a single fitted mixed model.

An advantage of this approach is that it can be implemented directly from a single fitted
mixed model, without requiring the analyst to construct a full FE or CRE analogue in settings
that depart from the standard individual-by-time panel, such as models with
multiple-membership random effects, nontrivial error-covariance structures $\bR$, or a
very large number of random-effect levels. In such applications, FE or CRE-style
specifications are often still possible in principle, but may require additional data
construction and modeling work. The value-added model in Section~\ref{sec:vam} provides
one such example. When conventional FE or CRE estimators are readily available, we
recommend using the classical Hausman or Mundlak--Wooldridge tests as global checks of
the RE specification, and then using the bias diagnostic to identify coefficients
or contrasts with pronounced fitted alignment, while keeping its interpretation
distinct from a numerical estimate of total bias.

\section{Gasoline Consumption Data Application}
\label{sec:gas}
We illustrate the calculations with the \texttt{Gasoline} data from
\texttt{plm}: 18 countries observed annually from 1960 through 1978
\citep{plm_gas,plm}. The response is log gasoline consumption per car;
the regressors are log income per capita, log gasoline price, and log
cars per capita. The analysis uses all 342 observations.

\par\addvspace{\baselineskip}
\noindent\begin{minipage}{\linewidth}
\begin{lstlisting}[caption={FE--RE comparison for the gasoline panel},label={lst1}]
library(plm)
data("Gasoline", package = "plm")
form <- lgaspcar ~ lincomep + lrpmg + lcarpcap
wi <- plm(form, data = Gasoline, model = "within")
re <- plm(form, data = Gasoline, model = "random",
          random.method = "swar")
phtest(wi, re)
\end{lstlisting}
\end{minipage}\par
The conventional Hausman calculation returns $H=302.8$. Its
covariance-difference matrix is indefinite, so we do not use the reported
$\chi^2$ tail probability as reliable evidence against the RE specification.
This example illustrates the covariance-estimation problem analyzed by
\citet[Section~2.2.1]{legallo2023}.

For the internal diagnostic, we fit the corresponding random-intercept
model by REML with \texttt{lme4} \citep{lme4}. This differs from the
Swamy--Arora covariance estimator used in Listing~\ref{lst1}.
The respective \texttt{plm} and \texttt{lmer} slopes are
$(0.5550,-0.4204,-0.6068)$ and $(0.5920,-0.3744,-0.6176)$.
The Hausman statistic and the diagnostics below therefore use different
fitted covariance weights and are not parts of one test. The calculations below use the prepared
\texttt{mixedbiastest} 1.1.0 revision; the accompanying reproduction
notes record the required software versions.

\par\addvspace{\baselineskip}
\noindent\begin{minipage}{\linewidth}
\begin{lstlisting}[caption={Seeded gasoline diagnostics and conditional simulations},label={lst2}]
library(lme4)
library(mixedbiastest)  # version 1.1.0
fit <- lmer(lgaspcar ~ lincomep + lrpmg + lcarpcap +
              (1 | country), data = Gasoline, REML = TRUE)
K <- list(lincomep = c(0,1,0,0), lrpmg = c(0,0,1,0),
          lcarpcap = c(0,0,0,1))
set.seed(20261022)
res <- mixedbiastest(fit, n_permutations = 100000,
                    k_list = K, verbose = FALSE)
print(res)
plot(res)
set.seed(20261023)
sim <- mixedbiastest::simulate_bias(fit, nsim = 2000)
sim$summary
\end{lstlisting}
\end{minipage}\par

\begin{table}[tbp]
\centering\small
\caption{Gasoline slopes: FE estimates, REML estimates, their difference,
the internal diagnostic $D_k$, and the permutation reference tail fraction.
The intercept is not separately estimable when all country effects are
treated as fixed, so the comparison is restricted to slopes.}
\label{tab:gas}
\input{gasoline_table.tex}
\end{table}

\begin{figure}[tbp]
\centering
\includegraphics[width=.94\textwidth]{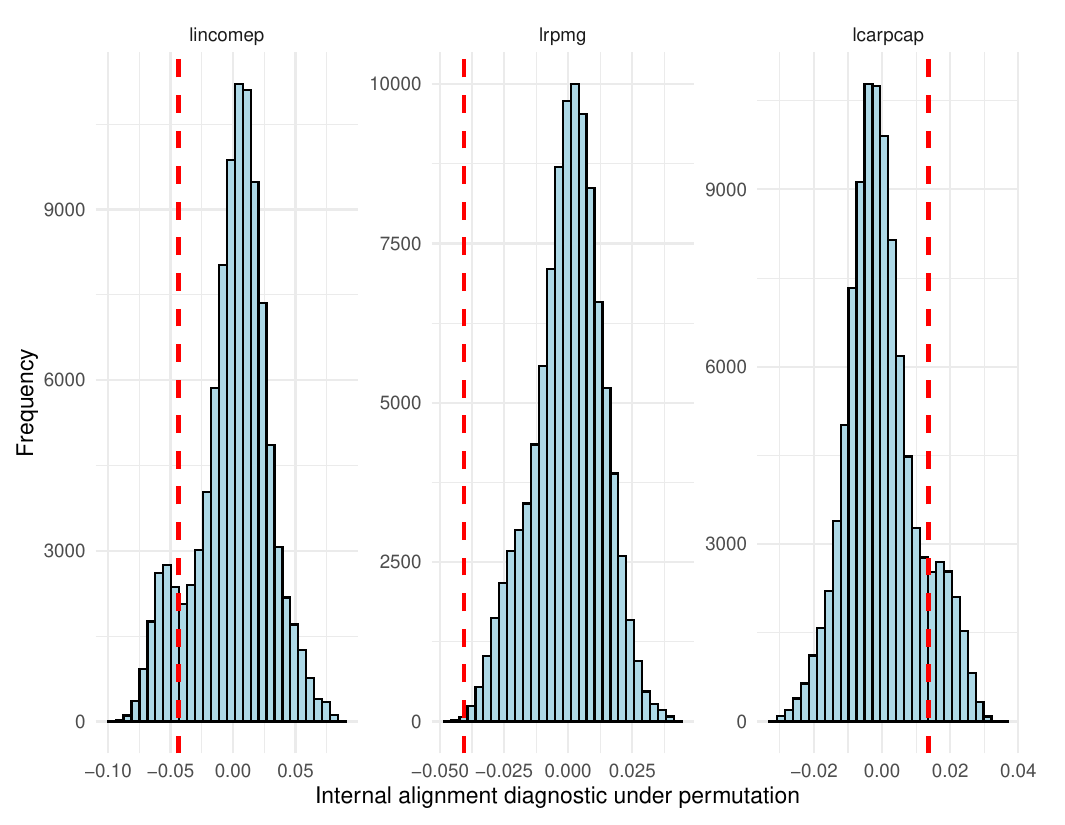}
\caption{Gasoline fitted-effect permutation references, based on 100,000
draws. Red dashed lines mark the observed internal diagnostics.
The histograms describe fitted alignment, not the sampling distribution
of the full coefficient error.}
\label{fig:gas}
\end{figure}

Table~\ref{tab:gas} and Figure~\ref{fig:gas} identify price as the clearest
alignment signal: its diagnostic is negative and lies in a tail of its
permutation reference. The tail fractions for income, price, and cars per
capita are approximately $0.1603$, $0.00062$, and $0.1982$, with Monte Carlo
standard errors $0.00116$, $0.000078$, and $0.00126$. These uncertainties
describe permutation sampling only. The signs agree with the RE-minus-FE
differences here, but the diagnostic values are appreciably smaller in
magnitude; neither comparison identifies the true bias in this panel.

\begin{table}[tbp]
\centering\small
\caption{Gasoline conditional response-refit simulation. The original
fitted effects and design are fixed; 2,000 independent residual draws
are generated and the REML covariance parameters are refitted. Parentheses give Monte
Carlo standard errors. The simulated diagnostic is recomputed in each
refit. Latent alignment plus weighted error equals total; the diagnostic
column is not part of that sum. The weighted-error and total columns use
the paired control variate described after \eqref{eq:simulation}.}
\label{tab:simulation}
\setlength{\tabcolsep}{4pt}
\input{gasoline_simulation_table.tex}
\end{table}

The generating model for Table~\ref{tab:simulation} uses the original
fitted coefficients and random effects, with residual variance
$0.00857319$; the original fitted random-intercept variance is
$0.09397076$. All 2,000 refits converged. The estimated weighted-error contributions
are small but distinguishable from zero at the simulation precision:
about $0.00126$, $0.00067$, and $-0.00043$ for the three slopes.
With covariance weights frozen at the original fit, the total conditional
bias in this plug-in scenario would equal the original diagnostic exactly.
After refitting, the totals differ.
For example, the original income diagnostic (Table~\ref{tab:gas}) is $-0.04355$, while its simulated
total is about $-0.04689$ under this generating scenario. Refitting the
covariance changes the latent-alignment contribution as well as introducing
a weighted-error mean. The magnitudes of the weighted-error means are about 1.6--2.0\% of
the respective model-based coefficient standard errors: nonzero, but small
on that scale. The diagnostics recomputed in the simulated samples also
differ from the total conditional biases; the magnitudes of their means are about
51\%, 77\%, and 46\% of those totals. These comparisons concern the fitted
generating scenario, not the unidentified bias in the population generating
the gasoline data.

\section{Value-added Model (VAM) Results with GPvam}
\label{sec:vam}
Value-added models are used to summarize teachers' associations with
student test scores \citep{asavam}. The complete persistence (CP) model
\citep{ballou,mariano10,karlem} gives a useful example beyond a simple
random-intercept panel. Teacher effects have separate variance components
by grade; within-student residual correlations are represented by
blocks of $\bR$. Current and prior teachers enter $\bZ$, producing a
multiple-membership structure \citep{browne01,lock07}.
Nonrandom assignment can make teacher effects dependent on the design,
creating sensitivity of fixed-effect estimates to the working RE
assumptions \citep{lock2007,kz}.

We retain the elementary-school example analyzed by \citet{karlcpm},
with 2,834 students, 304 teacher-by-year effects, and 6,236 observed scores
over three years. The original analysis divided the recorded scores
by two; estimates below use that same scale. Explicit ML fitting
reproduces the earlier estimates. This matters because the current
\texttt{GPvam} default is REML. The example uses \texttt{GPvam}'s
diagnostic \citep{gpvam}, which accommodates the multiple membership and
non-diagonal residual covariance outside the supported scope of
\texttt{mixedbiastest}. The prepared \texttt{GPvam} 3.4-0 revision also
implements response-refit simulations. The coefficient and permutation
results were originally generated with version 3.3-1; numerical checks
confirmed the same ML CP fit and diagnostic under the updated implementation.

\par\addvspace{\baselineskip}
\noindent\begin{minipage}{\linewidth}
\begin{lstlisting}[caption={CP model and a named contrast; individual-level input is required},label={lst3}]
library(GPvam)
# data must contain y, teacher, year, student,
# Race_Ethnicity and Gender from the original records.
data$y <- data$y / 2
fit.cp <- GPvam(data, persistence = "CP",
  fixed_effects = ~as.factor(Race_Ethnicity) +
    as.factor(year) + as.factor(Gender) + 0,
  REML = FALSE, hessian = FALSE, tol1 = 1e-9,
  max.iter.EM = 5000, verbose = FALSE)
stopifnot(fit.cp$converged)
set.seed(20261022)
all.effects <- bias.test.custom(fit.cp, n_perms = 99999,
                               plot = FALSE)
k <- setNames(numeric(length(coef(fit.cp))), names(coef(fit.cp)))
k["as.factor(Race_Ethnicity)White"] <- 1
k["as.factor(Race_Ethnicity)Hispanic"] <- -1
set.seed(20261023)
bias.test.custom(fit.cp, n_perms = 999999, k_vectors = k,
                 plot = FALSE)
\end{lstlisting}
\end{minipage}\par

\begin{table}[tbp]
\centering\small
\caption{CP model ML fixed-effect estimates, model-based standard errors,
internal diagnostics, and permutation reference tail fractions from 99,999
draws. Race coefficients are group intercepts; year 1 and female students
are the reference categories for the other covariates.}
\label{tab:cp}
\input{vam_table.tex}
\end{table}

The fitted alignment is negative for the Hispanic coefficient and
positive for the White and Asian/Pacific Island coefficients
(Table~\ref{tab:cp}). The White--Hispanic contrast estimate is $1.0847$
with model-based standard error $0.1031$ and internal diagnostic $0.1287$.
None of its 999,999 sampled permutations was as extreme, giving the
finite Monte Carlo value $10^{-6}$ under \eqref{eq:permutation}, rather
than zero. A REML sensitivity fit gives a contrast of $1.0836$ and a
diagnostic of $0.1284$.

We also applied the response-refit calculation to this ML CP fit, holding
the observed score pattern, covariates, and teacher assignments fixed and
using the original fitted coefficients and teacher predictions as generating
values. We drew correlated
within-student errors and refitted the covariance parameters in each of 100
simulated samples; all refits converged. For the White--Hispanic contrast,
the mean latent-effect contribution was $0.16497$
(Monte Carlo SE $0.00161$), and the weighted-error contribution
was $0.00009$ ($0.00068$), giving total
conditional bias $0.16506$ ($0.00161$).
The weighted-error and total estimates use the fixed-reference control variate described
after \eqref{eq:simulation}.

The weighted-error contribution is small on the scale of the original
contrast's model-based standard error, $0.1031$, in this conditional
simulation. An approximate 95\% Monte Carlo interval for this contribution
is $[-0.00125,0.00144]$, with both endpoints within 1.4\% of that standard
error. This interval reflects simulation error, not uncertainty about the
chosen generating scenario. The run does not resolve the contribution's
sign or establish that it is exactly zero.

With covariance weights frozen at the original fit, the total conditional
bias would equal the original diagnostic, $0.12871$, because the generating
teacher effects are the original fitted predictions. Refitting changes the
weights applied to that same fixed teacher signal, increasing its mean
contribution to $0.16497$. Thus, of the estimated $0.03635$ gap between the
original diagnostic and the total conditional bias, $0.03626$ comes from
the change in latent-effect weighting and only $0.00009$ from the
weighted-error term. Almost all of the estimated increase therefore comes
from refitting the covariance weights acting on the teacher signal.

The mean recomputed diagnostic was $0.09194$ (Monte Carlo SE $0.00144$).
It uses each refit's newly predicted teacher effects, whereas the latent
contribution uses the fixed generating teacher effects, so these quantities
need not agree. These calculations retain the shrinkage in the generating
teacher predictions and do not identify bias from the population assignment
mechanism. The supplied \texttt{vam\_refit\_simulation.R} records the two
independent random-number streams and reproduces the calculation.

The fitted diagnostics and conditional simulations warrant investigation,
but they are not causal effects of demographic characteristics. The
fitted-effect permutation reference need not be centered at zero. For scale, the original fitted diagnostic is about 12\% of the fitted contrast and 1.25
times its model-based standard error. Whether this sensitivity is practically important
depends on the intended use; the generating assignment mechanism remains unspecified.
The individual student records are not included with this revision;
the accompanying code records the required input schema and supplies
the aggregate results for inspection.

\section{Extensions and Future Directions}
\label{sec:future}
\subsection{Stochastic regressors and conditional bias}
The conventional panel setting need not be forced into a stochastic-assignment
interpretation. Let $\mathcal D=(\bX,\bZ)$ be a random design, assume
$\bE(\bds{\epsilon}\mid\mathcal D)=0$, and write
$\mu(\mathcal D)=\bE(\bse\mid\mathcal D)$. For a prespecified contrast $\bk$ and estimator $\hat{\bsB}=\bM_{\mathcal D}\by$,
where the GLS matrix $\bM_{\mathcal D}$
is specified given the design and $\bk'\bM_{\mathcal D}\bX=\bk'$, the conditional
bias is simply
\begin{equation}\label{eq:randomX}
\bE(\bk'\hat{\bsB}-\bk'\bsB\mid\mathcal D)
=\bk'\bM_{\mathcal D}\bZ\mu(\mathcal D).
\end{equation}
A fixed country-incidence matrix therefore does not preclude bias from
dependence of country effects on stochastic regressors. The conditional mean,
rather than whether the random matrix is called $\bX$ or $\bZ$, is the relevant
quantity, as in the conditional-mean formulation of correlated random effects
\citep{mundlak,wooldridge}. With response-fitted covariance weights, the conditional version of
\eqref{eq:bias} must again retain both expectations.

\subsection{From a local diagnostic to a coefficient path}
The diagnostic of \citet{kz} is the negative derivative of the fitted contrast
with respect to the log scale of the random-effects covariance; in balanced
random-intercept panels, this derivative reparameterizes the
sensitivity studied by \citet{vasnev2010}. To express the connection for the
mixed model in \eqref{eq:mixedmodel}, freeze the fitted covariance matrices
and vary a scalar multiplier $t>0$:
\begin{equation}\label{eq:path}
\begin{split}
\bV_t&=\hat{\bR}+t\bZ\hat{\bG}\bZ',\\
\hat{\bsB}(t)&=(\bX'\bV_t^{-1}\bX)^{-1}\bX'\bV_t^{-1}\by,\\
\hat{\bse}(t)&=t\hat{\bG}\bZ'\bV_t^{-1}
\{\by-\bX\hat{\bsB}(t)\}.
\end{split}
\end{equation}
Here $\bX$ is full column rank, $\hat{\bR}$ is positive definite, and
$\hat{\bG}$ is positive definite. Differentiating the GLS normal equations gives
\begin{equation}\label{eq:derivative}
-\frac{d\{\bk'\hat{\bsB}(t)\}}{d\log t}
=\bk'(\bX'\bV_t^{-1}\bX)^{-1}\bX'\bV_t^{-1}\bZ\hat{\bse}(t)
=:D_k(t).
\end{equation}
This is an identity for the observed data, including when the starting
covariances were estimated and then frozen. If the contrast is also estimable
after treating the random effects as fixed, its path converges to that
residual-weighted FE estimate as $t\to\infty$, and
\begin{equation}\label{eq:integral}
\bk'\hat{\bsB}(1)-\bk'\hat{\bsB}_{\mathrm{FE}}
=\int_1^\infty D_k(t)\,\frac{dt}{t}.
\end{equation}
Appendix~\ref{app:path} gives the derivation and the endpoint condition.
In balanced random-intercept panels the within estimator is already one end
of the estimator curve of \citet[Figure~4]{vasnev2010}.
For spherical residual covariance, the endpoint also follows from the
penalized-fixed-effects representation of \citet[Section~3.1]{hazlett2022}:
increasing $t$ weakens the penalty on the effects.
The diagnostic at $t=1$ describes local movement; the RE-minus-FE difference
accumulates movement over the entire path. They need not agree in magnitude,
and different parts of a path can cancel. This comparison is one concrete
way to examine features of a sensitivity curve beyond its derivative at a
single point, as suggested by \citet[p.~190]{magnusvasnev2007}.

\begin{figure}[tbp]
\centering
\includegraphics[width=\textwidth]{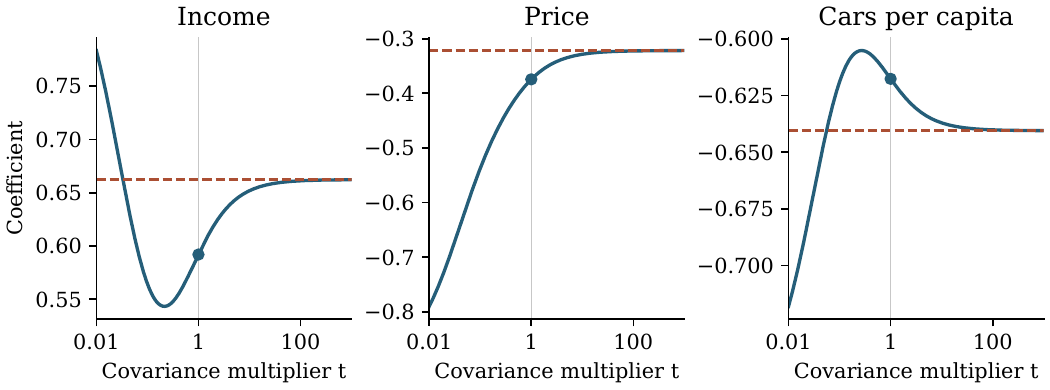}
\caption{Gasoline coefficient paths with the fitted residual and
random-effects covariance shapes held fixed. The dot is the fitted RE
coefficient ($t=1$); the dashed line is its country-FE endpoint.
These are descriptive paths, not calibrated confidence bands.}
\label{fig:paths}
\end{figure}
Figure~\ref{fig:paths} uses the same three slopes as Table~\ref{tab:gas}.
On the computed grid with $t\ge1$, all three paths move monotonically
toward their FE endpoints
and their diagnostics retain one sign, consistent with the matching signs in
Table~\ref{tab:gas}. In this example the RE-minus-FE gaps are 1.3--1.7 times
the diagnostics at $t=1$; the local slope does not summarize all movement
along the path. The reversals in the income and cars-per-capita paths occur
below $t=1$, on the side approaching pooled OLS as $t\to0$.

\subsection{What further inference would require}
Conditioning on stochastic regressors clarifies the source of bias, but does
not identify $\mu(\mathcal D)$ from one fitted model. In particular, a
conditional mean $\bZ\mu(\mathcal D)$ lying in the column space of $\bX$
can be indistinguishable from a change in $\bsB$ and invisible to every
covariance-scale diagnostic. A contrast that remains estimable when all
random effects are treated as fixed is unaffected by this particular
departure. Nor does a small diagnostic at one scale rule
out cancellation between random-effect directions. These limits motivate
examining the path as well as the diagnostic's value at the fitted covariance.

A further step is to state substantively plausible bounds on the
conditional mean departure and assess how coefficient conclusions change
over that class, in the spirit of omitted-variable sensitivity analysis
\citep{cinelli2020}. Such bounds could support sensitivity intervals rather
than an automatic bias adjustment; confidence intervals that account for
bounded bias are studied by \citet{armstrong2018}. Calibration of such intervals for the present mixed-model diagnostic with
estimated covariance parameters, especially near variance-component boundaries,
requires additional theory. Assessing performance over stochastic designs
also requires simulations that regenerate those designs. The conditional response-refit calculation
reported here is a useful component of that work, not a substitute for it.

\section{Conclusion}
The internal diagnostic gives a signed, coefficient-specific view of
random-effect alignment from one fitted mixed model. It complements the
Hausman or Mundlak--Wooldridge comparison when those specification tests are
appropriate, and remains computable in more complex multiple-membership
models. Its interpretation requires two distinctions: fitted alignment is
not total finite-sample bias, and a fitted-effect permutation reference is
not automatically a calibrated test.

A practical analysis can report the coefficient, its internal diagnostic,
and the assumptions behind any permutation or simulation comparison.
Response-refit simulations retain the extra weighted-error contribution
under a stated generating scenario, while covariance-scale paths show
whether the local diagnostic represents broader movement in the estimate.
Together these calculations make the sensitivity assessment more
informative without implying that the observed data identify a universal
bias correction.

\appendix
\section{Covariance-scale identity and FE endpoint}
\label{app:path}
Put $K=\bZ\hat{\bG}\bZ'$ and $N_t=\bX'\bV_t^{-1}\bX$.
Since $d\bV_t^{-1}/dt=-\bV_t^{-1}K\bV_t^{-1}$, differentiation of
$\bX'\bV_t^{-1}\{\by-\bX\hat{\bsB}(t)\}=0$ yields
\[
\frac{d\hat{\bsB}(t)}{dt}
=-N_t^{-1}\bX'\bV_t^{-1}K\bV_t^{-1}
\{\by-\bX\hat{\bsB}(t)\}.
\]
Multiplication by $-t\bk'$ proves \eqref{eq:derivative}. This calculation
uses the GLS fitted-value differential given by
\citet[proof of Theorem~2, p.~316]{banerjeemagnus1999}.
For the endpoint, use the penalized-fixed-effects representation of mixed
models; see \citet[Section~3.1]{hazlett2022} for the spherical-error case.
With the residual weighting used here, $\hat{\bsB}(t)$ and $\hat{\bse}(t)$ minimize
\[
(\by-\bX b-\bZ u)'\hat{\bR}^{-1}(\by-\bX b-\bZ u)
+t^{-1}u'\hat{\bG}^{-1}u.
\]
Let $(b_*,u_*)$ be an unpenalized minimizer and let $L_*$ be its residual
loss. The penalized solution has residual loss between $L_*$ and
$L_*+t^{-1}u_*'\hat{\bG}^{-1}u_*$. Its fitted mean therefore converges
to the unique $\hat{\bR}^{-1}$-weighted projection of $\by$ onto the
column space of $[\bX\ \bZ]$. If $\bk'b$ is estimable under this
augmented design, it is a linear functional of that fitted mean and
converges to the FE contrast. Integrating \eqref{eq:derivative}
from 1 to $T$ and taking $T\to\infty$ gives \eqref{eq:integral}
as an improper integral. Neither calculation asserts an exact sampling
distribution after covariance estimation.

\bibliographystyle{agsm}
\bibliography{disbib4}

\end{document}

%% file: gasoline_table.tex
\begin{tabular}{lrrrrr}
\toprule
Slope & FE & RE (REML) & RE $-$ FE & $D_k$ & $p_{\rm perm}$ \\
\midrule
Income & 0.6622 & 0.5920 & -0.0703 & -0.0435 & 0.16027 \\
Price & -0.3217 & -0.3744 & -0.0527 & -0.0405 & 0.00062 \\
Cars per capita & -0.6405 & -0.6176 & 0.0229 & 0.0136 & 0.19821 \\
\bottomrule
\end{tabular}

%% file: gasoline_simulation_table.tex
\begin{tabular}{lrrrr}
\toprule
Slope & \shortstack{Mean simulated\\diagnostic} & Latent alignment & Weighted error & \shortstack{Total conditional\\bias} \\
\midrule
Income & -0.02415 (0.00032) & -0.04815 (0.00009) & 0.00126 (0.00008) & -0.04689 (0.00010) \\
Price & -0.03558 (0.00010) & -0.04705 (0.00012) & 0.00067 (0.00003) & -0.04637 (0.00011) \\
Cars per capita & 0.00660 (0.00012) & 0.01490 (0.00002) & -0.00043 (0.00003) & 0.01447 (0.00003) \\
\bottomrule
\end{tabular}

%% file: vam_table.tex
\begin{tabular}{lrrrr}
\toprule
Coefficient & Estimate & Std.\ error & $D_k$ & $p_{\rm perm}$ \\
\midrule
Black & 24.2894 & 0.1612 & 0.0061 & 0.74535 \\
Am.\ Indian/Alaska & 24.3683 & 0.3228 & -0.0364 & 0.15388 \\
Asian/Pacific Island & 25.7910 & 0.1863 & 0.0506 & 0.01075 \\
Hispanic & 24.3989 & 0.1058 & -0.0691 & 0.00047 \\
White & 25.4836 & 0.0894 & 0.0597 & $10^{-5}$ \\
Year 2 & 0.9831 & 0.0600 & 0.0023 & 0.54544 \\
Year 3 & 1.9738 & 0.0876 & 0.0012 & 0.75405 \\
Male & 0.0552 & 0.0819 & 0.0036 & 0.49654 \\
\bottomrule
\end{tabular}